\documentclass[aps,pra,a4paperpaper,floatfix,showpacs]{revtex4} 
\usepackage{graphicx}
\usepackage{amssymb}
\usepackage{epstopdf}
\begin{document}
\title{Decoherence and the conditions for the classical control of quantum systems.}
\author{G. J. Milburn}
\affiliation{Centre for Engineered Quantum Systems, School of Mathematics and Physics, The University of Queensland, St Lucia QLD 4072, Australia. }

\begin{abstract}
We find the conditions for one quantum system to function as a classical controller of another quantum system: the controller must be an open system and rapidly diagonalised in the basis of the controller variable that is coupled to the controlled system. This causes decoherence in the controlled system that can be made small if the rate of diagonalisation is fast.  We give a detailed example based on the quantum optomechanical control of a mechanical resonator. The resulting equations are similar in structure to recently proposed models for consistently combining quantum and classical stochastic dynamics. 
\end{abstract}
\maketitle

\section{Introduction}
The results of measurements made on quantum systems are classical random variables extracted from classical instruments. A great deal of effort has been spent in trying to reconcile this fact with the apparent universality of quantum theory. In one approach, referred to as the decoherence program\cite{zurek}, the open, dissipative nature of a measurement apparatus becomes central.  An equivalent, although often ignored, problem of consistency arises when we ask for the conditions under which the control parameters in a quantum system can be regarded as entirely classical: why not simply include the control apparatus as part of the quantum system as well?  In this paper it will be shown that the  classical controllability of a quantum system also requires us to make explicit reference to the open and dissipative nature of control devices (lasers for example). 

The question of specifying the conditions under which we can treat the control of quantum systems using classical signals has arisen in a variety of fields. Most recently it has engendered some debate in the field of quantum information\cite{Gea-Banacloche, vanEnk}. Quantum computing requires one to be able to implement rather complex time dependent Hamiltonians for many interacting quantum system( qubits). If any of these subsystems became entangled with the control apparatus, it would not be possible to perform unitary control and errors would necessarily enter the quantum computation. For example, in ion trap quantum computing we do not concern ourselves with fundamental quantum nature of the controlling laser fields, or the RF trapping fields which are treated entirely classically. A very elegant discussion of this problem has been given by Carmichael and co workers\cite{carmichael1, carmichael2} who draw attention  to the fact that lasers are open quantum systems of a very special kind and the conditions for classical controllability turn out to be very similar to the conditions for the classical stochastic nature of measurement results. In the emerging field of hybrid engineered quantum systems we will need to consider a greater variety of quantum controllers than simply a laser.  

There is another control problem in which the control is intrinsically classical and cannot be treated as an appropriate limit of a quantum interaction; when the control is implemented by a gravitational field. This is related to the long debated inconsistencies that arise when we combine classical and quantum dynamics\cite{clas-quant}. One resolution has been proposed by Diosi\cite{Diosi}. It requires that we add noise to both the classical and the quantum dynamical system so that the noise spectral density in both cases are linked.  Adding noise to a classical dynamical system can be quite harmless, but adding noise to a quantum system necessarily implies decoherence.   

Wiseman and  Warszawski\cite{WW} discuss a related measurement problem that requires one to combine classical and quantum stochastic dynamics. They consider realistic measurement scenarios in quantum optics in which the actual observed stochastic process is conditioned not only on the quantum nature of the source but also by additional classical noise processes occurring in the detection circuits. The resulting equations are similar in structure to those obtained in this paper, as well as those postulated to provide a consistent hybrid classical and quantum dynamics.  A similar motivation was used by Reginatto and Hall \cite{RH} to consistently combine classical and quantum stochastic processes. An interesting application of their model was recently given by Chua et al. \cite{Chua}. A different kind of classical-quantum hybrid dynamics is discussed by Tsang\cite{Tsang} where the objective is to optimally estimate the classical, and possibly stochastic, controls on a quantum system by making continuous measurements on the quantum system.  The starting point of this paper is different from these models. Here we seek the conditions under which two interacting quantum systems can be regarded not as a measurement scenario but rather as a classical control scenario.   

A simple example will suffice to illustrate the relationship between classical controllability and the role of decoherence in measurement. Consider the case of applying a classical impulsive  force to a free particle,  that is to say,  a force applied on a  time scale much shorter than the time scale of the intrinsic dynamics of the particle. The change in the state of the system due to the impulsive force is well described by the unitary transformation $|\psi_{out}\rangle =U(X)|\psi_{in}\rangle$, where
\begin{equation}
U(X)=e^{-iX\hat{q}/\hbar}
\label{classical-kick}
\end{equation}
 where $X$ is a classical control parameter equal to the change in momentum of the particle due to the impulse and where $(\hat{q},\hat{p})$ are the quantum canonical position and momentum operators of the quantum system. 
 
 Now suppose the impulse was in fact due to a collision with another free particle --- call it the {\em control particle} for ease of reference --- with canonical position and momentum operators $(\hat{Q},\hat{P})$. In this case we know there will be a mutual exchange of momentum and the change in the state through the collision may be described by the unitary operator
 \begin{equation}
U=e^{-i\kappa \hat{Q}\hat{q}/\hbar}
\end{equation}
Under what conditions are we justified in describing this interaction by the unitary transformation in Eq.(\ref{classical-kick})? 

The answer of course depends on how we prepare the state of the control particle, $|\Phi\rangle_c$ for it is easy to see that the change in the state of the  target particle is given by
\begin{equation} 
\rho_{out}= \int_{-\infty}^\infty dX\  P(X) U(X) \rho_{in} U^\dagger(X)
\label{mixed-out}
\end{equation}
where 
\begin{equation}
P(X)= \frac{1}{\kappa}\left |\Phi_c\left (\frac{X}{\kappa}\right )\right |^2
\end{equation}
with $\Phi_c(Q)=\langle Q|\Phi\rangle_c$ is the position probability amplitude of the control particle immediately before the collision. In this form we see that we may regard the transformation as a mixture of classical controls with a random control parameter distributed according to $P(X)$. 

As an example, consider the case for which $|\Phi_c(Q)|^2$ is a Gaussian sharply peaked at a particular value $\bar{Q}$ with variance $\sigma$. We first change the variable of integration to $y=X-\bar{X}$, with where $\bar{X}=\kappa \bar{Q}$.  We can then factor out an overall unitary transformation, 
\begin{eqnarray*} 
\rho_{out} &= & U(\bar{X})\left [\int_{-\infty}^\infty dy\  (2\pi\sigma\kappa^2)^{-1/2}e^{-\frac{y^2}{2\sigma\kappa^2}}e^{-iy\hat{q}/\hbar} \rho_{in} e^{iy\hat{q}/\hbar}\right ]U^\dagger(\bar{X})
\end{eqnarray*}
If we  expand the unitary operators $e^{\pm iy\hat{q}/\hbar}$ we can integrate term by term. If $\frac{\sigma\kappa^2}{\hbar^2} << 1$, we can truncate this expansion to second order.   The output state in Eq.(\ref{mixed-out}) may be written as 
\begin{equation}
\rho_{out}= U(\bar{X})\rho_{in}U^\dagger(\bar{X})-\frac{\Gamma}{8}\left [\hat{q},\left [\hat{q}, U(\bar{X})\rho_{in}U^\dagger(\bar{X})\right ]\right ]
\end{equation}
 and $\Gamma=\frac{4\sigma\kappa^2}{\hbar^2}$. We see that provided $\Gamma <<1$, we can treat the control as entirely classical. The correction term adds noise to the momentum of the target particle, as one easily sees that
\begin{equation}
\langle \Delta \hat{p}^2\rangle_{out}=\langle \Delta \hat{p}^2\rangle_{in}+\frac{\hbar^2\Gamma}{4}
\end{equation}

The mixed state in Eq.(\ref{mixed-out}) begs an interpretation in terms of a measurement performed on the control. Suppose that, immediately after the impulsive interaction between the control and the target, the position of the control is measured with extreme accuracy.  The resulting conditional state of the target system, given a result $Q$ from the measurement, is given by  projecting the output state onto $|Q\rangle_c$ and tracing out the control. Thus
\begin{equation}
|\psi^{(Q)}\rangle_{out}=e^{-i\kappa Q\hat{q}/\hbar}|\psi\rangle_{in}=e^{-iX\hat{q}/\hbar}|\psi\rangle_{in}
\end{equation}
which indicates a classical control of the target subject to intrinsic quantum fluctuations of the control system position coordinate.  In order to make this small we need to ensure that $\sigma$ is small, i.e the position coordinate of the controller is well defined. 

It is worth contrasting this model with how one would use this kind of impulsive interaction to make a measurement of the system position variable, $\hat{q}$. We now regard the control particle not as a controller but as a measurement apparatus, a meter.  Clearly there is no point in measuring the position of the controller immediately after the impulse as this does not change. Instead one must measure the canonically conjugate momentum coordinate $\hat{P}$. For this to be a good measurement we need to prepare the state of the controller not with a well defined position but rather with a well defined momentum; exactly the opposite to what is required to use the system as a classical control. A quantum controller and a quantum meter are in this sense complementary devices. In the models we discuss below this kind of complementarity is determined by what controller variable is coupled to its environment.

\section{Quantum stochastic model.}
The impulsive model of the previous section is a pump to our intuition however it is rather limited. In practice we are more likely to be interested in a system evolving continuously in time subject to a classical controller. In this section we show that we can still regard the classical controller as a quantum system subject to continuous measurement of the controller variable coupled to the system. This of course means we need to treat the quantum controller as an open quantum system. It is the nature of this interaction that enables us to describe the target system dynamics in terms of a purely classical stochastic control. 

Consider two interacting quantum systems, a {\em controller} and a {\em target} (see figure \ref{fig1}). The control system is also coupled to an environment in such a way that one particular degree of freedom in the environment may be monitored provding a continuous weak measurement of the controller variable that is coupled to the target system.  The Hamiltonian describing the dynamics is 
\begin{equation}
H=H_c+H_s+\hbar\kappa \hat{X}\hat{x}+H_e+H_{ce}
\end{equation}
where $\hat{X}$ is a (dimensionless) hermitian operator for the control and $\hat{x}$ is a (dimensionless) Hermitian operator on the system. We will assume that $\hat{X}$ has a continuous spectrum. The Hamiltonians $H_e$ and $H_{ce}$ represent, respectively,  the Hamiltonian for the environment coupled to the control and the interaction Hamiltonian for the control and its environment. We will assume that $H_{ce}$  is simply linear in $\hat{X}$ so that in principle the environment can be regarded as a measurement apparatus for the control variable $\hat{X}$. What we are interested in is the dynamics of the target system alone, given the interaction with the control system and given the entire measurement record of the control variable $\hat{X}$. 

The theory of weak continuous measurement is now well established\cite{WM}. In the case of measurement of the control system observable $\hat{X}$ we use the theory described in \cite{CCM}. There are two ways to view the measurement which we might loosely refer to as the inside view and the outside view. The first of these simply gives the unconditional dynamics of the measured system when no account is taken of the measured results.  The dynamics is given in terms of a master equation for the measured system density operator by averaging over the state of the environment. The inside view gives the stochastic conditional dynamics of the measured system conditioned on a particular measurement record.The conditional dynamics is given by a nonlinear stochastic master equation and is sometimes called a quantum trajectory. The reason the dynamics is non linear is to ensure that the future measurement records are conditioned on what was seen in the past. 

We will take the unconditional dynamics to be given by 
\begin{equation}
\frac{d\rho}{dt} = -\frac{i}{\hbar}[H_c+H_s, \rho]-i\kappa [\hat{X}\hat{x},\rho]-\frac{\Gamma}{4}[\hat{X},[\hat{X},\rho]]
\label{unconditional}
\end{equation}
where $\Gamma$ is the {\em decoherence rate} and determines the rate of decay of the off diagonal matrix elements of $\rho$ in the basis of the measured quantity. It simultaneously adds noise to the observable canonically conjugate to $\hat{X}$.   The corresponding conditional dynamics is given by the Ito stochastic master equation,
\begin{equation}
\label{conditional_me}
d\rho^c = -\frac{i}{\hbar}[H_c+H_s, \rho^c]dt-i\kappa [\hat{X}\hat{x},\rho^c]-\frac{\Gamma}{4}[\hat{X},[\hat{X},\rho^c]]dt+\sqrt{\frac{\Gamma}{2}}{\cal H}[\hat{X}]\rho^c dW(t)
\end{equation}
where
\begin{equation}
{\cal H}[A]\rho = A\rho+\rho A^\dagger -{\rm tr}(A\rho+\rho A^\dagger)\rho
\label{con_operator}
\end{equation}
and $dW(t)$ is the Wiener stochastic process\cite{GZ} and the superscript on the state serves to remind us that this state is conditional on the entire measuremtent record up to time $t$. The classical stochastic process corresponding to the observed measurement record is,
\begin{equation}
dy(t) = \Gamma \langle \hat{X}\rangle^c+\sqrt{\frac{\Gamma}{2}}dW(t)
\end{equation}
where $\langle \hat{X}\rangle^c={\rm tr}( \hat{X} \rho^c)$ is the conditional mean.  The decoherence rate, $\Gamma$ has a dual role.  It  determines both the decoherence rate and the signal to noise ratio. Indeed $\Gamma >>1$ implies rapid decoherence and large signal to noise ratio, which we simply refer to as the good measurement limit. In what follows we set $H_c=0$ so that the control variable has no dynamics other than what is implied by the continuous measurement made upon it. We will remove this assumption in the next section when we consider simultaneous measurement of canonically conjugate observables on the control system. 

The conditional stochastic master equation tends to drive the system to eigenstates of the measured quantity. To see this let us suppose for simplicity that the coupling to the target is turned off $\kappa\rightarrow 0$ and that we ignore the free dynamics of the control (that is to say, we assume it is very slow compared to the measurement dynamics). The uncondtiional dynamics of the controller is 
\begin{equation}
\label{control-uncond}
\frac{d\rho_c}{dt} = -\frac{\Gamma}{4}[\hat{X},[\hat{X},\rho_c]]
\end{equation}
and the conditional dynamics is 
\begin{equation}
\label{control-cond}
d\rho_c^c =-\frac{\Gamma}{4}[\hat{X},[\hat{X},\rho_c^c]]dt+\sqrt{\frac{\Gamma}{2}}{\cal H}[\hat{X}]\rho_c^c dW(t)
\end{equation}

In the good measurement limit the unconditional state will be rapidly diagonalised in the eigenstates of $\hat{X}$. We can thus assume that  the unconditional state for the controller, on time scales of interest, can be written as 
\begin{equation}
\rho_c(t)=\int_{-\infty}^\infty dX P(X,t)|X\rangle_c\langle X|
\end{equation}
Substituting this into the unconditional master equation, Eq.(\ref{control-uncond}) we see that it implies $d\rho_c(t)=0$ and thus $P(X,t)=P(X,0)$. On the other hand substituting this into the conditional master equation, Eq.(\ref{control-cond}), we find that the solution has the same form but that the {\em conditional}  distribution $P^c(X,t)$ obeys the classical conditional dynamics 
\begin{equation}
dP^c(X,t) = \sqrt{2\Gamma}(X-\langle X\rangle^c)P^c(X,t)dW
\label{class-cond}
\end{equation}
In this sense the control system, in the good measurement limit, may be described as a classical stochastic process subject to continuous observation. Note that averaging over the noise, we have that $dP(X,t)=0$, which is consistent with the unconditional dynamics $d\rho_c(t)=0$ and $P(x,t)=P(x,0)$. This simply says that the diagonal matrix elements of the controller, in the basis that diagonalises the measured quantity, do not change.  If we further assume that the initial state has a Gaussian Wigner function, the conditional dynamics will remain Gaussian and we can derive the following equations for the conditional mean $\langle X\rangle^c={\rm tr}(\hat{X}\rho^c_c) $ and conditional variance of $\sigma ={\rm tr}( \Delta \hat{X}^2 \rho^c_c)$. Taking care with the Ito calculus, we find that the conditional mean and conditional variance satisfy,
\begin{eqnarray}
d\langle X\rangle^c & = & \sqrt{2\Gamma}\sigma dW(t)\\
d\sigma & = & -2\Gamma \sigma dt
\label{stochastic-eqns}
\end{eqnarray}
The conditional mean is stochastic but the conditional variance is deterministic and decays exponentially to zero at the decoherence rate. In other words, for a good measurement,  the conditional state will localise on a random eigenstate of the measured quantity.

Turning now to the unconditional dynamics including the target system, we make the ansatz that the solution is well approximated, on time scales of interest to the target dynamics, by 
\begin{equation}
\rho(t)=\int_{-\infty}^\infty dX\ \rho_s(X,t)\otimes|X\rangle_c\langle X|
\end{equation}
We further assume that the controller dynamics is little changed from what it would be in the absence of the coupling to the target. This assumes that $\Gamma$ is much larger than all the system frequencies. It also accords with our intuition for a classical controller, that is to say, a classical controller should not be affected by the system it is used to control. Taking the partial trace of this state over the target system gives the state of the control. Consistency then requires that we set 
\begin{equation}
{\rm tr}_s\rho_s(X,t)=P(X,t)
\end{equation}
We can then write the unconditonal state as 
\begin{equation}
\rho(t)=\int_{-\infty}^\infty dX\ P(X,t) \rho_s(t|X)\otimes|X\rangle_c\langle X|
\label{ansatz}
\end{equation}
where
\begin{equation}
\rho_s(t|X)=\frac{\rho_s(X,t)}{P(X,t)}
\end{equation}
The unconditional state of the target system alone is obtained by taking the partial trace of $\rho(t)$ in Eq.(\ref{ansatz})
\begin{equation}
\rho_s(t) =\int_{-\infty}^\infty dX\ P(X,t) \rho_s(t|X)
\end{equation} 
Substituting this into the unconditional master equation we find that
\begin{equation}
d\rho_s(t) = -\frac{i}{\hbar}[H_s,\rho_s(t)]dt-i\kappa[\hat{x},\hat{M}_s(t)]dt
\end{equation}
where
\begin{equation}
\hat{M}_s(t)=\int_{-\infty}^\infty\ dx  XP(X,t) \rho_s(t|X)
\end{equation}
To simplify this equation we use $P(X,t)=P(X,0)$ and approximate
\begin{eqnarray}
\rho_s(t) & \approx &  \rho_s(t|X_0)+\frac{\sigma_0}{2}\left .\frac{\partial^2\rho(t|X)}{\partial X^2}\right |_{X_0}\\
\hat{M}_s(t) & \approx & X_0\rho(t|X_0)+\sigma_0 \left .\frac{\partial \rho(t|X)}{\partial X}\right |_{X_0}
\end{eqnarray}  
where $X_0, \sigma_0 $ are the mean and variance of $P(X,0)$ respectively.  We now proceed iteratively in powers of $\sigma$. We first use the zeroth order solution $\rho_s(t)=\rho_s(t|X_0)$ to get 
\begin{equation}
d\rho_s(t|X_0) = -\frac{i}{\hbar}[H_s,\rho_s(t|X_0)]dt-i\kappa X_0[\hat{x},\rho_s(t|X_0)]dt
\end{equation}
which may be solved as 
\begin{equation}
\rho(t|X_0) = \exp\left [-\frac{i}{\hbar} H_s t-i\kappa X_0 t\hat{x}\right ]\rho_s(0)\left [\frac{i}{\hbar} H_s t+i\kappa X_0 t\hat{x}\right ]
\end{equation}
which in turn gives the next order approximation, 
\begin{equation}
d\rho_s(t|X_0) = -\frac{i}{\hbar}[H_s,\rho_s(t|X_0)]dt-i\kappa X_0[\hat{x},\rho_s(t|X_0)]dt-\kappa^2\sigma_0[\hat{x},[\hat{x},\rho(t|X_0)]]
\end{equation}
We can pause here to note an important feature already present in the simple model of the introduction. The double commutator in $\hat{x}$ will cause decoherence in the position basis and drive fluctuations into the momentum of the controlled system. We get classical control provided we can make this term small enough over the time scales of interest. 
Provided that $\sigma\kappa^2$, which has units of frequency, is much smaller than the inverse time scale of the system, we can disregard the decoherence and take the dynamics to be very closely approximated by the Hamiltonian $H=H_s+\hbar\kappa X_0\hat{x}$ with $X_0$ a  classical control parameter.

On the other hand, given that we have assumed a measurement of the controller,  we do have access to the actual measurement record and can use this to describe the conditional dynamics of the target system.  We now postulate that the conditional state of the total system is given by
\begin{equation}
\rho^c(t)=\int_{-\infty}^\infty dX\ P^c(X,t) \rho^c_s(t|X)\otimes|X\rangle_c\langle X|
\label{cond-ansatz}
\end{equation}
where $P^c(X,t)$ satisfies the classical conditional equation, Eq.(\ref{class-cond}). It is easier to interpret this equation if we 
write $P^c(X,t)$ in functional form
\begin{equation}
P^c(X,t)=\int dX_t \delta(X-X_t){\cal P}^c(X_t)
\label{stochastic-path}
\end{equation}
where   ${\cal P}^c(X_t)$ is a distribution over stochastic paths. We can write the conditional  state of the combined control and target system as 
\begin{equation}
\rho_s(t)=\int  dX_t\ {\cal P}^c(X_t) \rho_s(X_t)\otimes|X_t\rangle_c\langle X_t|
\end{equation}
where to ease notation we have set $\rho_s(t|X_t)\equiv \rho(X_t)$. 
It is then a simple matter to see that
\begin{equation}
d\rho^c_s(X_t)=-\frac{i}{\hbar}[H_s,\rho^c_s(X_t)]-i\kappa X_t[\hat{x},\rho^c_s(X_t)]
\end{equation}
where  $X_t$ is the observed classical control process in Eq.(\ref{class-cond}). In the example of a Gaussian process this is given by the equations Eq.(\ref{stochastic-eqns}).  In the case of a very good measurement $X_t$ will settle down to a constant with vanishing fluctuations. 

 The key feature of this model is the rapid decoherence of the controller in the basis of the control variable acting on the system ($\hat{X}$). This occurs because the classical controller was assumed to be continuously monitored in this basis. In the language of decoherence, the eigenstates of $\hat{X}$ are the {\em pointer basis} for the measurement. In general the pointer basis is determined by how the controller is coupled to its environment.  In order to make a quantum system behave as a classical controller we must engineer it so that the required control variable that is to act on the system is the pointer variable for controller.  In the simple example of this section the system-controller interaction Hamiltonian commutes with the pointer basis variable of the controller. 
 
 It is interesting to contrast this result with the role of the pointer basis in a continuous measurement model\cite{WM}. In the quantum theory of continuous measurement, a quantum system can play the role of the first stage of a measuring instrument, let us call it the meter, provided that the meter variable that is coupled to the system does {\em not commute} with the pointer basis variable.  In the example of this section, we could use the controller as a meter by choosing the interaction Hamiltonian to be of the form $\hbar\kappa\hat{P}\hat{x}$ where $\hat{P}$ is canonically conjugate to the pointer basis variable $\hat{X}$. In this case a continuous measurement of $\hat{X}$ will give information on the system variable $\hat{x}$.  A quantum system engineered to function as a classical controller can be regarded as {\em complementary} to a quantum system engineered to be a measurement apparatus. In both cases a key role is played by the pointer basis of the controller/meter. In other words it is how a quantum system is coupled to its environment that fixes its use as either a classical controller or as a measurement apparatus.

\section{An optomechanical example.}
In this section we give a less idealised example for how to engineer a classical controller as an open quantum system. The example is taken from the field of quantum optomechanics in which the mechanical action of the electromagnetic field can be used to control or measure the motion of a mechanical resonator.   In particular we will consider the case of a single mode optical cavity field interacting with a mechanical resonator comprising one mirror of the cavity\cite{OM}. 

The interaction between the field and the mechanics is known as the radiation pressure interaction as it corresponds to a force applied to the mechanical element that is simply proportional to the photon number inside the cavity.  The Hamiltonian is given by\cite{OM}
\begin{equation}
H_{om}=\hbar\omega_c a^\dagger a +\frac{\hat{p}^2}{2m}+\frac{m\omega_m^2}{2}\hat{x}^2 +\hbar(\omega_c/L) a^\dagger a\hat{x}
\end{equation}
where $a,a^\dagger$ are the annihilation and creation operators for the single mode cavity field, $\hat{x},\hat{p} $ are the canonical displacement and momentum operators for the mechanical resonator and  $\omega_c$ is the frequency of the optical cavity mode, $\omega_m$ is the mechanical frequency of the moving mirror with effective mass $m$ and $L$ is the length of the optical cavity. It will be more convenient to write the displacement and momentum of the mechanical element in terms of the lowering ($b$) and raising operators ($b^\dagger$)
\begin{eqnarray}
\hat{x} & = & \left (\frac{\hbar}{2m\omega_m}\right)^{1/2}(b+b^\dagger) \\
\hat{p} & = & -i\sqrt{\frac{\hbar m\omega_m}{2}}(b-b^\dagger)
\end{eqnarray}
Then,
\begin{equation}
H_{om}=\hbar\omega_c a^\dagger a +\hbar \omega_m b^\dagger b +\hbar G_0 a^\dagger a(b+b^\dagger)
\label{OM_ham}
\end{equation}
where 
\begin{equation}
G_0=\frac{\omega_c}{L} \left (\frac{\hbar}{2m\omega_m}\right)^{1/2}
\end{equation}
is known as the optomechanical vacuum coupling rate. For most systems currently in operation $G_0$ varies from a  few Hz \cite{Aspelmeyer} to a fractions of a MHz \cite{Painter}.  We will regard the cavity field as the controller and mechanical resonator as the system to be controlled.  

At first sight it would appear from the Hamiltonian in Eq. (\ref{OM_ham}) that the classical control variable of interest is the photon number inside the cavity and the required pointer variable is the photon number. However this is misleading as we need first to consider how the cavity is coupled to its environment. The (approximate) pointer basis variable for a damped cavity mode is not the intracavity photon number but rather the amplitude of the field in the cavity\cite{WallsMilb2}. We thus need to drive the cavity in such a way that a steady state field can build up. The small position dependent detuning that results from the radiation pressure interaction is then turned into a small displacement of the cavity field amplitude which does define the pointer basis for the damped cavity. 

The quantum theory of a single damped cavity mode is well known\cite{WM,GZ} and is described by the master equation
\begin{equation}
\left (\frac{d\rho}{dt}\right )_{cav}=-i\omega_c[ a^\dagger a,\rho]+\kappa{\cal D}[a]\rho
\end{equation}
where
\begin{equation}
{\cal  D}[A]\rho = A\rho A^\dagger -\frac{1}{2}\left (A^\dagger A \rho+\rho A^\dagger A\right )
\end{equation}
and $\kappa$ is the decay rate of the photon number inside the cavity. 
The pointer basis states for this master equation are the coherent states $|\alpha\rangle$ defined by $a|\alpha\rangle=\alpha|\alpha\rangle$\cite{WM,WallsMilb2}. In the absence of any coherent driving of the cavity, the steady state is simply the vacuum state $|0\rangle$ which is a coherent state. More interesting is the case of a cavity driven by a coherent laser source with carrier frequency $\omega_L$. The steady state of the cavity in this case is a coherent state with coherent amplitude $\alpha_0$ given by
\begin{equation}
\alpha_0=\frac{-i\sqrt{\kappa}{\cal E}_L}{\kappa/2+i\Delta}
\label{ss_amp}
\end{equation}
where $\Delta=\omega_c-\omega_L$ is the detuning of the driving laser from the cavity and units have been chosen so that $|{\cal E}_L|^2$ are in units of photon flux (i.e. number per second).  This result can be obtained by including in the master equation a hamiltonian term of the form
\begin{equation}
H_d=\hbar\sqrt{\kappa}\left (a^\dagger {\cal E}_L  e^{-i\omega_L t}+a {\cal E}_L^*  e^{i\omega_L t}\right )
\label{driving}
\end{equation}

At this point we note that we are describing the laser by a classical control parameter and this might be regarded as question begging.  It looks like we need to go back one stage to describe the laser itself as a classical control field. This reasoning will lead to an infinite regress very much like the infinite regress that results if we insist on regarding every stage of a measurement as a quantum system. We need to put a quantum-classical cut somewhere and in the case of measurement this is determined by how certain key degrees of freedom of the measurement apparatus are coupled to the environment. However we do not face this problem here.  As discussed in some detail by Noh and Carmichael\cite{carmichael2}, the description of laser driving used here is good approximation given the particular kind of open irreversible system that is a laser. Their explanation parallels the view taken in this paper. We will not consider this further here as our objective is to determine under what conditions we can regard an optomechanical interaction as a classical controller for the mechanical resonator. 

If we now include the driving Hamiltonian and the damping of the cavity mode, the dynamics of the optomechanical system, in an interaction picture with respect to the cavity field at frequency $\omega_L$ is given by
\begin{equation}
\frac{d\rho}{dt} =-\frac{i}{\hbar}[H_I,\rho]+\kappa {\cal D}[a]\rho
\label{om_me}
\end{equation}
where  
\begin{equation}
H_I=\hbar \Delta a^\dagger a +\hbar\omega_m b^\dagger b+\hbar G_0 a^\dagger a (b+b^ \dagger)+\hbar(E^*a+Ea^\dagger)
\end{equation}
and $E=\sqrt{\kappa}{\cal E}_L$ which now has units of frequency. 

We now follow the approach of the previous section; we first consider the steady state of the controller itself and then use this to get approximate equations of motion, including fluctuations, for the target system under the assumption that the irreversible dynamics of the controller is much faster than the target system dynamics. We then consider the conditional dynamics given that the state of the controller is continuously monitored. In the case of the optomechanical model this will be done via heterodyne detection which is a means of monitoring the cavity field amplitude. 

The steady state of the master equation for the controller, in the absence of the coupling to the target system, is a coherent state given by
\begin{equation}
\rho_c=|\alpha_0\rangle\langle \alpha_0|
\end{equation}
where $\alpha_0$ is given in Eq.(\ref{ss_amp}). In the presence of the coupling to the mechanical system, we assume that the total unconditional state may be written as
\begin{equation}
\rho(t)=\int d^2\alpha \rho_b(t,\alpha)|\alpha \rangle_a\langle \alpha|
\label{ansatz_om}
\end{equation}
Taking the partial trace over the controller (i.e. the cavity) gives the unconditional state of the mechanical resonator as
\begin{equation}
\rho_b(t)=\int d^2\alpha \rho_b(t,\alpha)
\end{equation}
Taking the partial trace over the mechanical resonator gives the  unconditional state of the controller as
\begin{equation}
\rho_a(t) = \int d^2\alpha P(t,\alpha)|\alpha\rangle_a\langle \alpha|
\label{cavity-state}
\end{equation}
where 
\begin{equation}
P(t,\alpha)={\rm tr}\rho_b(t,\alpha)
\end{equation}
The form of Eq.(\ref{cavity-state}) implies that we are defining a Glauber-Sudarshan representation for the cavity state. We may need to admit some unusual distributions for this function (for example, derivatives of delta-functions) in order to account for non-classical effects. Note that $\rho_b(t,\alpha)$ is not a normalised state of the mechanical resonator. We may define a normalised unconditional state for the mechanical resonator as 
\begin{equation}
\rho_b(t|\alpha)=\frac{\rho_b(t,\alpha)}{P(t,\alpha)}
\end{equation}
and we have written the normalised state {\em as if} it were a conditional state conditioned on a random variable $\alpha$, with distribution $P(t, \alpha)$. However this assumes that $P(t,\alpha)$ is a regular probability density which may not always be the case.  

Substituting the assumption, Eq.(\ref{ansatz_om}), into the master equation, and using the correspondence rules for the P-representation\cite{WallsMilb},  we find that
\begin{eqnarray}
\label{partial-me}
\frac{\partial \rho_b(t,\alpha)}{dt} & = & -i\omega_m[b^\dagger b,\rho_b(t,\alpha)]-iG_0|\alpha|^2[b+b^\dagger,\rho_b(t,\alpha)]\\\nonumber
& & +\frac{\partial}{\partial \alpha}\left ((i\Delta+\kappa/2)+iE\right )\rho_b(t,\alpha)+\frac{\partial}{\partial \alpha^*}\left ((-i\Delta+\kappa/2)-iE\right )\rho_b(t,\alpha)\\\nonumber
& & +iG_0\frac{\partial}{\partial \alpha} \alpha(b+b^\dagger)\rho_b(t,\alpha)-iG_0\frac{\partial}{\partial \alpha^*} \alpha^*\rho_b(t,\alpha)(b+b^\dagger)
\end{eqnarray}
If we take the trace of this equation over the mechanical resonator states, we find that
\begin{eqnarray}
\label{FP-eqn}
\frac{\partial P(t,\alpha)}{\partial t} & = & \frac{\partial }{\partial \alpha}\left ( (i\Delta+\kappa/2)+iE\right )P(t,\alpha)+\frac{\partial }{\partial \alpha^*}\left ( (-i\Delta+\kappa/2)-iE\right )P(t,\alpha)\\\nonumber
& & +\left [\frac{\partial }{\partial \alpha}\left ( iG_0\alpha \langle b+b^\dagger\rangle^{(\alpha)}\right ) +\frac{\partial }{\partial \alpha}\left (-iG_0\alpha^* \langle b+b^\dagger\rangle^{(\alpha)}\right )\right ]P(t,\alpha)
\end{eqnarray}
where we have defined the {\em conditional} mean value
\begin{equation}
\langle b+b^\dagger\rangle^{(\alpha)}={\rm tr}\left ((b+b^\dagger)\rho_b(t|\alpha)\right )
\end{equation}
The form of Eq.(\ref{FP-eqn}) is almost the classical Liouville for the controller density $P(t,\alpha)$, but the conditional mean in the last term gives this a more complicated dynamics, as this term needs to be found by also solving Eq.(\ref{partial-me}). The classical and quantum dynamical system are tied together in a particular way. We shall return to this interpretation in section \ref{discussion} when we discuss the connection to work on hybrid quantum-classical dynamics.

We can now find the master equation for the unconditional state of the mechanical resonator, the controlled system, by integrating both sides of  Eq.(\ref{partial-me}) over the complex variable $\alpha$. This takes a rather simple form after integrating by parts, 
\begin{equation} 
\label{me-control}
\frac{d\rho_b(t)}{dt}=-i\omega_m[b^\dagger b,\rho_b(t)]-iG_0\left [b+b^\dagger, \int d^2\alpha |\alpha|^2 P(t,\alpha)\rho_b(t|\alpha)\right ]
\end{equation}
This again begs an obvious interpretation: it looks like the system is evolving with a random Hamiltonian. However the random variable itself is {\em not} independent as it is tied to the system through the last term in Eq. (\ref{FP-eqn}). 

In order to proceed we need to make some assumptions about the rate of dissipation in the controller, i.e. the cavity field damping rate, $\kappa$. 
We now assume that the cavity is strongly damped so that $\kappa>> \Delta, G_0$. Under this assumption we can expand the total system state near the cavity vacuum state as 
\begin{eqnarray}
\label{ad_ansatz}
\rho(t)& = & \int d^2\alpha  \left [\rho_{00}(t,\alpha)|\alpha\rangle_a\langle \alpha|+\rho_{01}(t,\alpha)|\alpha\rangle_a\langle \alpha|(a-\alpha)+\rho_{10}(t,\alpha)(a^\dagger-\alpha^*)|\alpha \rangle_a\langle \alpha|\right .\\\nonumber
& &\, \, \,  \, \, \,   \, \, \,  \, \, \,  \, \, \,   \, \, \,  \left .+\rho_{11}(t,\alpha)(a^\dagger-\alpha^*)|\alpha\rangle_a\langle \alpha|(a-\alpha)\right ]  \delta^{(2)} (\alpha-\alpha_0)\\\nonumber
&  = & \rho_{00}(t,\alpha_0)|\alpha_0\rangle_a\langle \alpha_0|+\rho_{01}(t,\alpha_0)|\alpha_0\rangle_a\langle \alpha_0|(a-\alpha_0)+\rho_{10}(t,\alpha_0)(a^\dagger-\alpha_0^*)|\alpha_0 \rangle_a\langle \alpha_0|\\\nonumber
& & \, \, \,  \, \, \,   \, \, \,  \, \, \,  \, \, \,   \, \, \, +\rho_{11}(t,\alpha_0)(a^\dagger-\alpha_0^*)|\alpha_0\rangle_a\langle \alpha_0|(a-\alpha_0)
\end{eqnarray}
where the operators $\rho_{jk}(t, \alpha)$ act only on the Hilbert space of the mechanical system.  This can be written in the form given in Eq.(\ref{ansatz_om}) using known correspondences\cite{WallsMilb} for the P-representation of the raising and lowering operators on the cavity field. We find that
\begin{eqnarray}
\label{total-P-rep}
\rho(t) & = & \int d^2\alpha |\alpha\rangle_a\langle \alpha|\left [\left (\rho_{00}(t,\alpha)+\rho_{11}(t,\alpha)\right )\delta^{(2)}(\alpha-\alpha_0)\right .\\\nonumber
& & \, \, \, \, \, \, \, \, \, -\left (\frac{\partial}{\partial\alpha^*}\rho_{01}(t,\alpha)+\frac{\partial}{\partial\alpha}\rho_{10}(t,\alpha)\right )\delta^{(2)}(\alpha-\alpha_0)\\\nonumber
& & \, \, \, \, \, \, \, \, \, \left .+\frac{\partial^2}{\partial\alpha^*\partial\alpha}\ \rho_{11}(t,\alpha)\delta^{(2)}(\alpha-\alpha_0)\right ]
\end{eqnarray}
The state of the mechanical resonator is then obtained by tracing out the control in Eq.(\ref{ad_ansatz}) to get
\begin{equation}
\rho_b=\rho_{00}(t,\alpha_0)+\rho_{11}(t,\alpha_0)
\end{equation}
If we substitute the expression in the integrand on the right hand side of  Eq.(\ref{total-P-rep}) into Eq.(\ref{me-control}) and integrate by parts we find that
\begin{eqnarray}
\label{approx-mech-me}
\frac{d\rho_b(t)}{dt} & = & -i\omega_m[b^\dagger b,\rho_b(t)]-iG_0|\alpha_0|^2\left [b+b^\dagger, \rho_{00}(t,\alpha_0)+\rho_{11}(t,\alpha_0)  \right ]\\\nonumber
& & -iG_0\left [b+b^\dagger, \alpha_0\rho_{01}(t,\alpha_0)+\alpha_0^*\rho_{10}(t,\alpha)+\rho_{11}(t,\alpha)\right ]
\end{eqnarray}

On the other hand, substituting the approximation in Eq.(\ref{ad_ansatz}) into the master equation, Eq.(\ref{om_me}), and equating coefficients we obtain equations for $\frac{d\rho_{jk}}{dt}$. These show that the off-diagonal terms rapidly approach a steady state (i.e. $\dot{\rho}_{10}=\dot{\rho}_{01}=0$), and we can  solve for the off-diagonal terms in terms of the diagonal terms, 
\begin{eqnarray}
\rho_{10} & = &  -\frac{iG_0\alpha_0}{\kappa/2+i\Delta}((b+b^\dagger)\rho_{00}-\rho_{11}(b+b^\dagger))\\
\rho_{01} & = & \frac{iG_0\alpha_0}{\kappa/2-i\Delta}(\rho_{00}(b+b^\dagger)-(b+b^\dagger)\rho_{11})
\end{eqnarray}
This procedure is equivalent to the adiabatic elimination of the cavity field as described in \cite{wiseman1993}. 
Substituting these into Eq.({\ref{approx-mech-me}) we find that to lowest order in $1/\kappa$, 
 \begin{equation}
 \label{reduced_me}
 \frac{d\rho_m}{dt}=-\frac{i}{\hbar}[H_m,\rho_m]+\Gamma{\cal D}[b+b^\dagger]\rho_m
 \end{equation}
 where 
 \begin{equation}
 \Gamma=\frac{4 G_0^2\alpha_0^2}{\kappa}
 \end{equation}
 and $H_m=\hbar\omega_m b^\dagger b+\hbar G_0|\alpha_0|^2(b+b^\dagger)$.  The target system is then seen to behave like a classically driven harmonic oscillator with driving determined by replacing $a^\dagger a\rightarrow |\alpha_0|^2 $ and a decoherence term in $\hat{x}$. This decoherence term tends to diagonalise the target system in the eigenbasis of $\hat{x}$ and simultaneously adds noise to the conjugate variable $\hat{p}$. This is the same result that we obtained for the simple model of the previous section. Provided that $\kappa >> \omega_m, (G_0\alpha_0)^2$ we can neglect this decoherence and treat the control as entirely classical. We again see that because the control system, the cavity field, is an open system with a particular pointer basis (the coherent states) we can find a limit in which the control may be treated classically.  
 
 To complete the discussion we turn now to the conditional dynamics that results when we continuously monitor the cavity field. A continuous measurement of the coherent field in a damped driven cavity requires a knowledge of two real numbers and can be done by heterodyne detection\cite{WM}. In essence this is a continuous simultaneous measurement of the two canonically conjugate  quadrature phase operators of the cavity field.   The resulting measurement record is described by two real stochastic processes, which may be combined to form a single complex stochastic process. The conditional master equation for a single damped cavity mode subject to heterodyne detection is 
\begin{equation}
\left (d\rho^c \right )_{cav}=-i\omega_c[ a^\dagger a,\rho]dt-i[E^*a+E a^\dagger,\rho^c]dt+\kappa{\cal D}[a]\rho^c dt +\sqrt{\kappa}{\cal H}[a]\rho^c dZ^*
\end{equation}
 where  the conditioning superoperator, ${\cal H}$,  is given in Eq.(\ref{con_operator}), and the complex valued Wiener process  $dZ=(dW_1+idW_2)/\sqrt{2}$ and
 \begin{eqnarray}
 \overline{dW_{i,j}} & = & 0\\
 \overline{dW_{i,j}^2}  & = & dt
 \end{eqnarray}
The classical measurement record is given by the in-phase and quadrature-phase heterodyne currents, $J_x,J_y$ which obey the classical stochastic (Ito) differential equations
\begin{eqnarray}
dJ_x(t) & = & \kappa\langle a+a^\dagger\rangle  dt+ \sqrt{2\kappa}dW_1\\
dJ_y(t)dt & = & -i\kappa\langle a-a^\dagger\rangle  dt+ \sqrt{2\kappa}dW_2
\end{eqnarray}
which may be more conveniently written in terms of the complex stochastic current, $J=(J_x+iJ_y)/2$, as
\begin{equation}
dJ(t)=\kappa\langle a\rangle^{c}(t)+\sqrt{\kappa}dZ
\end{equation}
Note that the superoperator, ${\cal H}$ gives zero when acting on coherent states; another indication that the coherent states are the pointer states for the damped cavity field.

Given that we do have access to the measurement results we can use them to classically control the mechanical resonator. We now show that this is consistent with the irreversible, unconditional dynamics of the controlled system we have already described. With this in mind we consider the stochastic Hamiltonian
\begin{equation}
H(t)=\hbar\omega_m b^\dagger b+\hbar G_0 \frac{|J(t)|^2}{\kappa}(b+b^\dagger)
\end{equation}
over a small time step $dt$ the system then evolves as
\begin{equation} 
\rho_b(t+dt)= dU \rho_b(t)dU^\dagger
\end{equation}
where 
\begin{equation}
dU= \exp\left [-\frac{i}{\hbar}dH_I(t)\right ]
\end{equation}

If we expand $dH_I(t)$ out to second order in the stochastic increment, taking care to keep the Ito increment $dZdZ^*$, and linearise around the steady state value $\langle a\rangle^{c}=\alpha_0$ we find the effective stochastic increment of the unitary evolution operator is 
\begin{equation}
dU(t)= \exp\left [-i\omega_m b^\dagger b dt-iG_0|\alpha_0|^2(b+b^\dagger) dt-i\frac{2G_0\alpha_0}{\sqrt{\kappa}}dW_1\right ]
\end{equation}
Expanding this to second order in the stochastic increment and averaging over the noise we find that the unconditional dynamics is given by the master equation in Eq. (\ref{reduced_me}). This indicates that it is entirely consistent to describe the dynamics conditionally using a classical stochastic control Hamiltonian based on the appropriate measurement record and the unconditional dynamics that results when we treat the open systems dynamics of the total system by adiabatically eliminating the controller. This is illustrated schematically in figure \ref{fig1}. 
\begin{figure}[h] 
   \centering
   \includegraphics[scale=0.5]{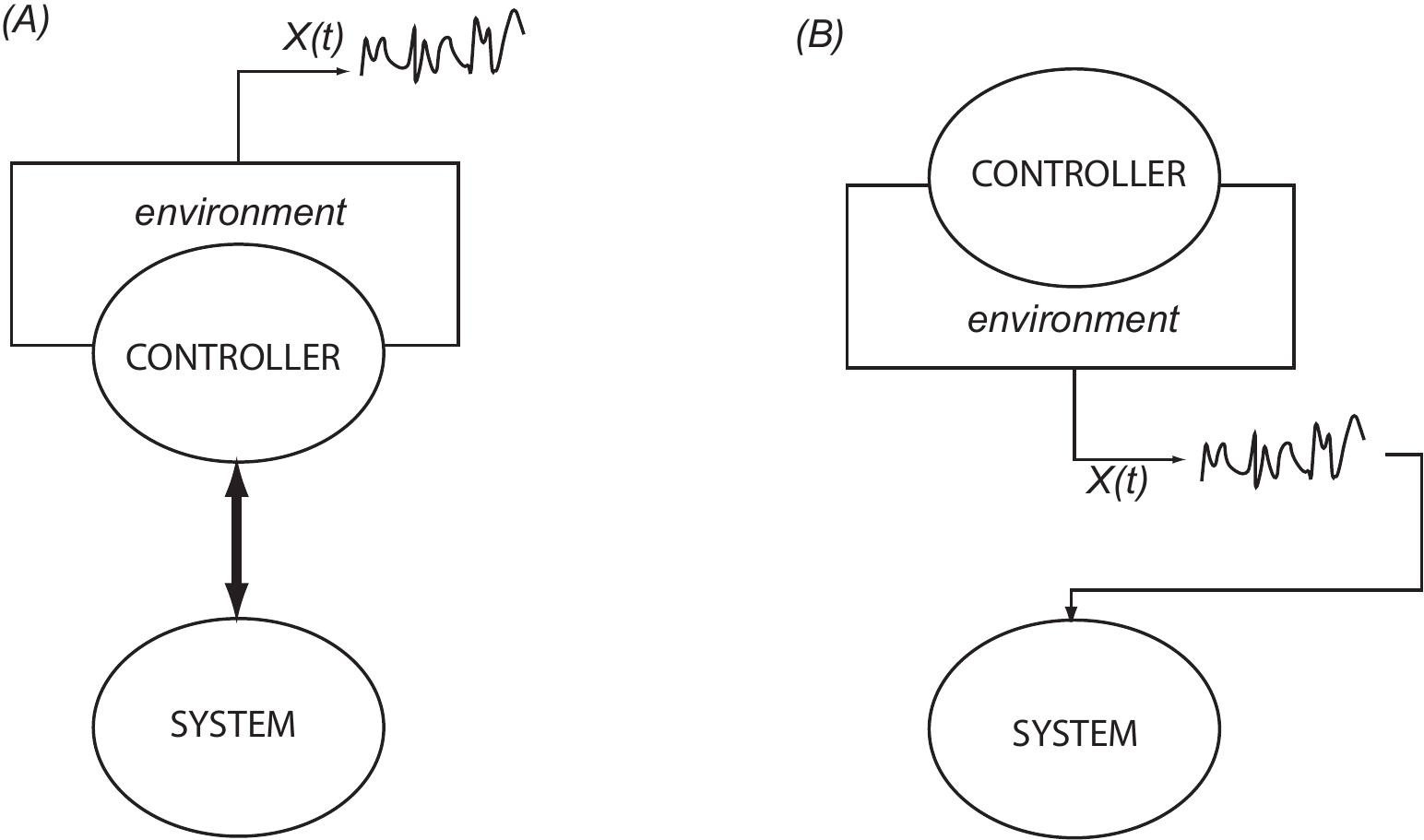} 
   \caption{Equivalent ways of describing the interaction between two quantum systems as classical control. In (A), the controller and the system interact unitarily but the controller is an open system subject to continuous measurement, with measurement record $X(t)$. In (B) the measurement record is used as a classical control signal to directly drive the target system. In both cases the unconditional  dynamics (average over the measurement record) and conditional dynamics (conditional on the measurement record) of the target system is the same.} 
   \label{fig1}
\end{figure}

\section{Discussion and conclusion}\label{discussion}
We have given two examples of interacting quantum systems in which one of the systems can be regarded as a classical controller for the other. The conditions necessary for this are, firstly, that the controller is open in such a way as to define a pointer basis for the controller operator that is coupled to the controlled system and secondly that the open system dynamics of the controller is such that the approach to the pointer basis is very rapid compared to the timescales of interest for the controlled system. In the optomechanical example we saw that the key features are captured by two linked equations, Eq.(\ref{partial-me}) and Eq.(\ref{FP-eqn}).  One equation gives the unconditional dynamics of the controlled quantum system alone in terms of an equation for a state operator $\rho(t,\alpha)$ that depends on a classical phase space variable, $\alpha$. The other equation gives the evolution for the classical distribution function $P(\alpha,t)$ that depends on a conditional mean of the controlled system variable occurring in the interaction $\langle b+b^\dagger\rangle^{(\alpha)}$. This structure parallels closely an approach to consistently combining quantum and classical dynamics developed by Diosi and others (see \cite{Diosi} and references therein). 

In \cite{Diosi} a classical system with canonical variables $(q,p)$ is interacting with a quantum system. The description begins by defining an operator $\rho(t, q,p)$ acting on the quantum system Hilbert space such that the actual state of the quantum system is given by
\begin{equation}
\rho_Q(t)=\int dqdp\rho(t, q,p)
\end{equation}
The operator $\rho(q,p)$ is called the {\em hybrid state}. The corresponding classical Liouville distribution for the classical system is defined by
\begin{equation}
P_c(t, q,p)={\rm tr}\rho(t, q,p)
\end{equation}
A hybrid observable is then defined as an operator on the quantum state space as $A(q,p)$ with average 
\begin{equation}
\langle A(t)\rangle =\int dqdp\ {\rm tr}\left [A(q,p)\rho(t, q,p)\right ]
\end{equation}

This is all well and good but do the dynamics correspond to anything that would make physical sense? Diosi shows\cite{Diosi} that it does provided the quantum system undergoes decoherence in just the right way and the classical system suffers a corresponding noise contribution. To see how this works we define a hybrid Hamiltonian $\hat{H}$,
\begin{equation}
\hat{H}=\hat{H}_Q+H_C(q,p)+\hat{H}_{QC}(q,p)
\end{equation}
where the last term is a hybrid observable intended to describe an interaction between the quantum and classical subsystems. The dynamics is then give by the quantum Liouville equation for the quantum system with an additional decoherence term. The classical evolution is given by a classical Fokker-Planck equation that includes a classical drift term via a noise averaged Aleksandrov bracket and  an additional classical diffusion term. The quantum decoherence rate and the classical diffusion rate are related so as to ensure the overall dynamics is a positive stochastic process.  In the optomechanical example the positivity is ensured by the construction as both the classical noise and the quantum decoherence ultimately originate from a positive quantum evolution equation for the controller and the target system. 

The hybrid dynamics defined by Diosi and others was motivated by the problem of reconciling classical gravitation with quantum dynamics. We have seen in this paper that similar models arise when a quantum system is used as a classical controller for another target quantum system. The possibility of classical control emerges as a result of the interaction of the controller with its environment. This is very similar to the way in which a measurement device works: a classical stochastic output is possible only if the measurement apparatus is coupled to the environment in this right way (i.e. define a pointer basis). It would be interesting to see if a similar possibility were open for gravity. Could the apparent classical nature of gravity be telling us something important about the nature of the underlying quantum theory? Perhaps the classical gravitational degrees of freedom are coupled to hidden environmental degrees of freedom in just the right way to make general relativity possible. If this were true there would be an unavoidable decoherence in the dynamics of every quantum system moving in a gravitational field. The real question however is, how big is the effect?

\begin{acknowledgments}
I gratefully acknowledge the support of the Australian Research Council Centre of Excellence for Engineered Quantum Systems.
\end{acknowledgments}

\end{document}